# Direct observation of electronic domains in manganites by spatially resolved spectroscopy

D.D. Sarma, Dinesh Topwal and U. Manju

*Solid State and Structural Chemistry Unit, Indian Institute of Science, Bangalore 560012, India*

S.R. Krishnakumar.

*International Centre for Theoretical Physics (ICTP), Strada Costiera 11, 34100 Trieste, Italy*

M. Bertolo, S. La Rosa and G. Cautero.

*Sincrotrone Trieste, S.S. 14 km 163.5 - in AREA Science Park, 34012 Basovizza - Trieste, Italy*

T. Y. Koo, P. A. Sharma and S-W. Cheong.

*Department of Physics & Astronomy, Rutgers University, New Jersey 08854, U.S.A.*

A. Fujimori.

*Department of Complexity Science and Engineering, University of Tokyo,*

*5-1-5 Kashiwanoha, Kashiwa, Chiba 277-8561, Japan*

The science of manganites is characterised by a close interplay between charge, spin and lattice degrees of freedom [1-3], arising from the similarities in the energy scales associated with each of these. Besides giving rise to spectacular phenomena [2], such as the colossal magnetoresistance, charge ordering, electronic and magnetic transitions induced by a variety of impulses (composition, temperature, magnetic field and photon field, to name a few), it has been recently realised that phase coexistence plays a very important role in controlling the properties of these materials. It is believed that many of these manganites contain an intricate mixture of different phases, typically a ferromagnetic metallic phase along with an antiferromagnetic insulating phase, at a sub-micron level. These have been probed both structurally using transmission electron microscope (TEM) [4] and magnetically using Lorentz force TEM techniques [5] and magnetic force microscopy [6]. Interestingly enough, there is so far no microscopic investigation that directly addresses the issue of metallic or insulating behaviour of these observed textures, though this phenomenon is known as the electronic phase separation, implying an inhomogeneous distribution of electrons in the matrix. It is widely held that the ferromagnetic domains are metallic, while the antiferromagnetic ones are insulating. However, the recent suggestion of the possibility of a ferromagnetic insulating phase opens up further possibilities. The main impediment in probing the electronic structure at a sub micron length scale has been to achieve sufficiently high energy resolution required to probe the electronic structure along with the ability to change the temperature of the sample over a wide range in a set-up that will also deliver a sufficiently intense photon source focussed on the sample at a sub micron size.

Besides the question concerning the electronic structure, there are several other issues that are not fully understood in the context of the science of phase separation. It is still unclear what really triggers this phase separation, with possibilities of underlying chemical inhomogeneities and strain effects being responsible. Theoretical models [7] have shown signatures of electronic phase separations even in absence of any such underlying inhomogeneity arising from composition fluctuations or strain fields. Recent developments in the synchrotron and associated instrumentations at the third generation synchrotron source, Elettra, has made it now possible to study the electronic structure by carrying out high resolution (< 50 meV) photoemission study with a spatial resolution of the order of 0.5 micron and where it is also possible to vary the temperature of the sample from down to about 40 K up to the room temperature using a continuous flow liquid He cryostat. We show a schematic of the complete set-up in Fig. 1. The focussing of the photon flux to a point source on to the sample is achieved using a Schwarzschild objective [8]. We have taken a large number of high-resolution spectra over a wide energy range from various points on the sample surface. We have also generated intensity images by recording the spectral intensities over the valence band region with the help of a sixteen-channel detector array. Since these samples do not cleave well, it is not possible to have the entire surface under investigation at the same level of focus for the electron detection system. Consequently, the absolute intensity at a given energy depends on the local topography, the height and the angle, of the probed part on the sample. We use the ratio of the intensity close to the Fermi level and that at the most intense valence band feature at about 3 eV below $E_F$ to partially cancel such topographic contributions. Since the intensity at $E_F$ is expected to be the highest in the highly metallic state and absent or nearly absent for the insulating state, the mapping of the ratio of intensities at these two energies, not only removes the topographic features to some extent, but also accentuates the distinction between the metallic and insulating phases. Essentially, these "ratio" images then represent metallic parts with a large value of the ratio and insulating parts with a small value of the ratio. We note that it is impossible to remove signatures of topographic variations in the intensity pattern completely; however, a comparison of a topographic image using the total signal intensity over all the sixteen channels and a ratio image clearly establishes which of the features arises from topography and which from a true variation in the electronic structure.

The experiments were carried out after cleaving single crystals of $La_{0.25}Pr_{0.375}Ca_{0.375}MnO_3$; these samples do not have any easy cleaving plane and therefore such cleaves invariably produce surfaces that are not atomically flat, giving rise to topographic variations. A large number of independent experiments were carried out to confirm the reproducibility of the results. At the two extreme temperatures (about 40 and 300 K) available at this set-up, the sample temperature could be kept stable for a long time, so the images were acquired with 100 x 100 pixels at both these temperatures; however, we recorded the image with coarser pixels (50 x 50) at all other intermediate temperatures in order to minimise the time to record one image and the consequent temperature drift during a single measurement. The vacuum in the chamber during the measurement was typically $1 \times 10^{-10}$ mbar. The metal-insulator transition in this sample has been established by monitoring the spatially averaged spectral intensity at the Fermi energy [9]; the spectrum at the low temperature shows the clear emergence of an appreciable signal intensity at the $E_F$ with a characteristic Fermi-Dirac statistics controlled spectral shape, establishing an overall metallic nature. At an elevated temperature, this feature is absent, suggesting an insulating state.

In order to ensure the absence of any significant level of chemical inhomogeneity that mars the experimental results, we have taken a large number of highly resolved spectra over a wide range of energy covering the main valence band consisting primarily of Mn 3d and O 2p states and also covering the energy range including La 5p, O 2s, Pr 5p and Ca 3p core levels; these spectra were collected over a freshly cleaved crystal face of size approximately $65 \times 32$ $\mu m^2$ and each spectrum is recorded with a spatial resolution of 0.5 micron. We have plotted these spectra with a single normalisation to make the highest intensity valence band feature at about 3 eV binding energy to have the same intensity in all spectra in order to account for topographic differences; we have also removed a linear background from the spectra of the shallow core levels to account for slightly changing background in some of the cases. Fig. 2a shows the collection of the valence band spectra and Fig. 2b those from the core level region with 30 spectra in all plotted in each of the panels. It is clear that the spectral features overlap on each other exceedingly well establishing that there is no chemical inhomogeneity at a length scale of 0.5 micron or larger in these samples.

Now we present the electronic structure images from freshly cleaved crystal faces in Fig. 3. The topmost single panel shows the topographic image with the total intensity of all the 16 channels of the detector added together. The panels immediately below show the ratio images, the highest ratio of the intensity at $E_F$ to that at the valence band maximum is denoted by the deep blue and the lowest intensity by the deep red. All the images show the same place on the crystal face at various temperatures. The stability of the particular part of the sample in view can be easily checked by the appearance of a slight topographic feature in the form of a semicircular part on the right part covering most of the ratio images and most prominently in the topographic image (topmost panel), distinct from the left part that has a lower overall intensity due to the crystal face being slightly out of focus there or tilted. Clearly, the first panel of the ratio image from the sample at 52 K shows a predominantly metallic nature by the colours ranging from green to deep blue; however, there is clearly an insulating piece signified by the red-yellow in the middle of the image, which we have surrounded by the thin black line for easy identification. Comparing with the topographic image (topmost panel), it is clear that this feature does not arise from any topographic peculiarity, but is a true reflection of a changed electronic structure. The spectrum within this red-yellow region is clearly that of the insulating phase with low intensity at $E_F$ as illustrated in Fig. 2a. This is a clear and direct evidence of the existence of a rather large insulating patch deep within the metallic phase, thereby establishing an electronic phase separation in this material. It is interesting to note that the colour contrast vanishes when the sample is warmed up, as shown in the second panel (253 K), thereby showing that the entire sample transforms in to the insulating phase at the elevated temperature. Most interestingly, when the sample is again cooled down to the lowest temperature (50 K), the colour contrast reappears, showing a metallic sample over most of the face, but with the insulating patch at about the same place as before. This is very significant for two reasons. First of all, the fact that the insulating patch in the first and the third panels are not quite exactly the same asserts that this is not a case of static phase separation with an insulating phase or impurity fixed in space, in which case precisely the same region of the sample space must reappear. The changes in the topology of the insulating patch provide evidence to support a dynamic reappearance of the insulating patch within the low temperature metallic phase. However, clearly there is an obvious memory effect and the patch reappears at about the same region in space with similar orientation and topology. This can only happen if there is something underlying that pins the nucleation sites of the insulating patch. Our results (Fig. 2) however establish that this pinning of the nucleation sites cannot be due to an underlying chemical inhomogeneity, giving credence to the speculation that strain fields may be responsible.

We show the evolution of the microscopic electronic structure of this crystal through a series of ratio images recorded at various temperatures between 50 and 90 K. It is clearly seen that the contrast in the electronic structure evident at the lowest temperature gradually vanishes due to the macroscopic metal-insulator transition with raising the temperature, closely following the metal-insulator transition observed in transport measurements. In order to make sure that these observations are not unique to the particular spot on the sample, we took similar images at the lowest temperature at various places on the crystal surface and in every case we found evidences of such electronic phase separation in terms of the existence of insulating patches deep inside the metallic regime. We illustrate this in the lowest three image panels in Fig. 3, recorded as the sample was cycled from the lowest temperature up to about 132 K and then down to the lowest T once again. In the first panel, we see an abundance of insulating patches, represented by red-yellow regions, marked by thin solid lines, in the image that predominantly represents a metallic phase, represented by blue-green regions. When the sample is heated up to about 132 K, the electronic phase contrasts vanish, as can be seen in the middle panel. However, the insulating patches reappear at approximately the same locations when the sample is cycled back to the lowest T, supporting the observations made above.

In conclusion, we have used a spatially resolved, direct spectroscopic probe for electronic structure with an additional unique sensitivity to chemical composition to investigate high quality single crystal samples of $La_{0.25}Pr_{0.375}Ca_{0.375}MnO_3$. The formation of distinct electronic domains has been observed in absence of any perceptible chemical inhomogeneity, where the relevant length-scale is at least an order of magnitude larger than all previous estimates. We have also provided compelling evidence, based on memory effects in the domain morphology, that electronic domain formation is intimately connected with long-range strains, often thought to be an important ingredient in the physics of this effect.

**FIGURE CAPTIONS**

Fig. 1
A schematic of the experimental set-up. The synchrotron beam is energy selected and focussed on to the sample in a small spot using the Schwarzschild Objective. The sample is mounted on a piezo drive, allowing one to scan the sample surface under the photon beam. At each position of the sample surface, a complete photoemission spectrum with high energy resolution can be recorded using a 16 channel electron analyser, providing high spatial and energy resolutions at the same time. The sample temperature can be varied between 40 and 300 K.

Fig. 2
Typical photoemission spectra obtained from 30 different spots on the sample surface with high energy and spatial resolution. (a) shows the valence band region and (b) shows the shallow core level region, covering all the elements present in the compound. The excellent overlap of all the spectra on top of each other suggests that there is not appreciable chemical inhomogeneity in the sample over the length-scale comparable with the spatial resolution of this set-up. In the case of the spectra in panel (b), linear backgrounds were subtracted from the spectra to account for slight variations in the background function contributed from different spots on the sample.

Fig. 3
The black-and-white images (size = $25 \times 54$ $\mu m^2$) show the topographic features of the sample surface at two different positions in terms of the total intensity collected at the analyser integrating over a large part of the valence band spectrum. The colour plots show a series of ratio images, as explained in the text, illustrating the microscopic changes in the spatially-resolved electronic structure on the sample surface as the sample is cycled through a temperature series. The black outlines drawn on the colour images provide a guide to the eye, encompassing the regions where we see signatures of electronic domain formation (insulating patches in an otherwise metallic state) at the low temperature and illustrating the memory effect.

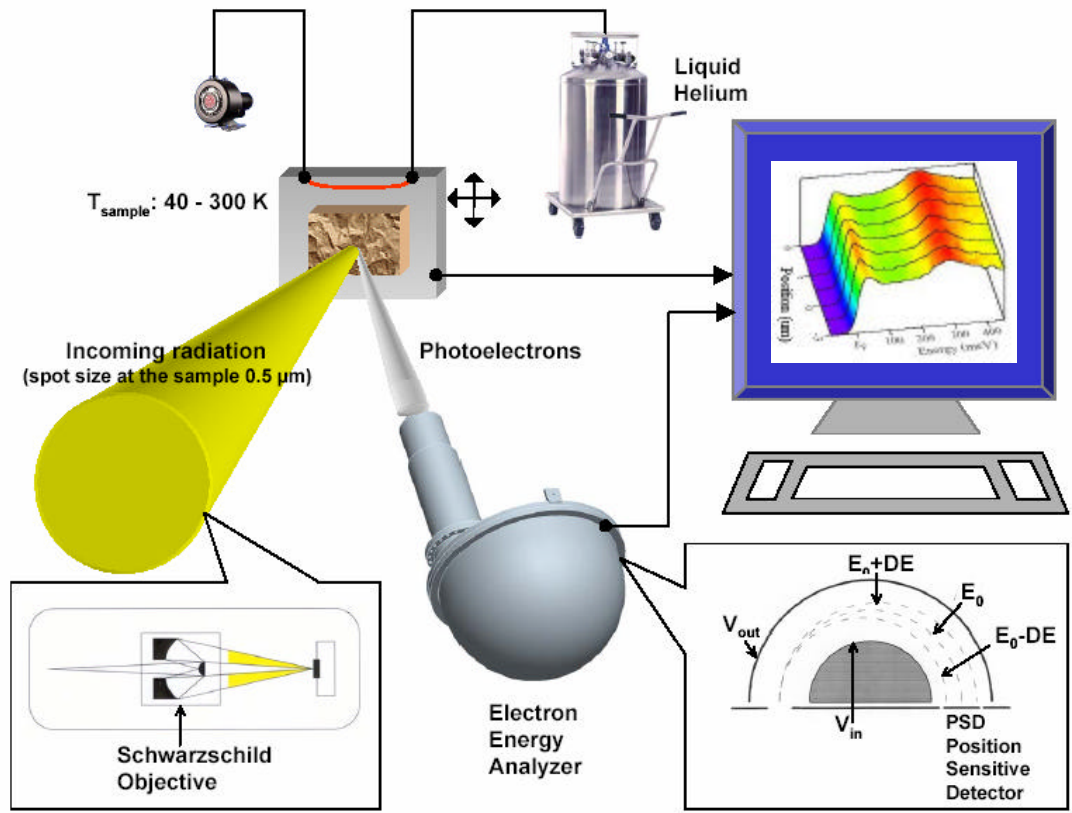

Fig 1 Sarma *et al*.

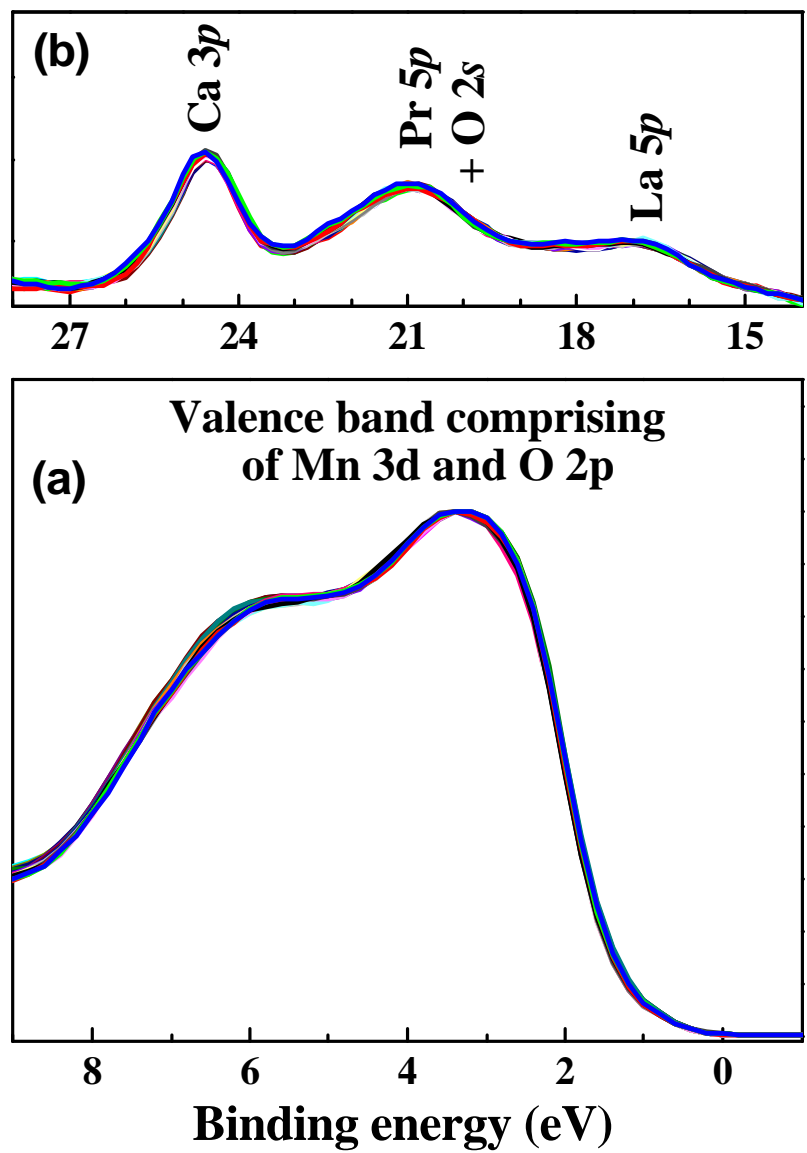

Fig 2 Sarma *et al.*

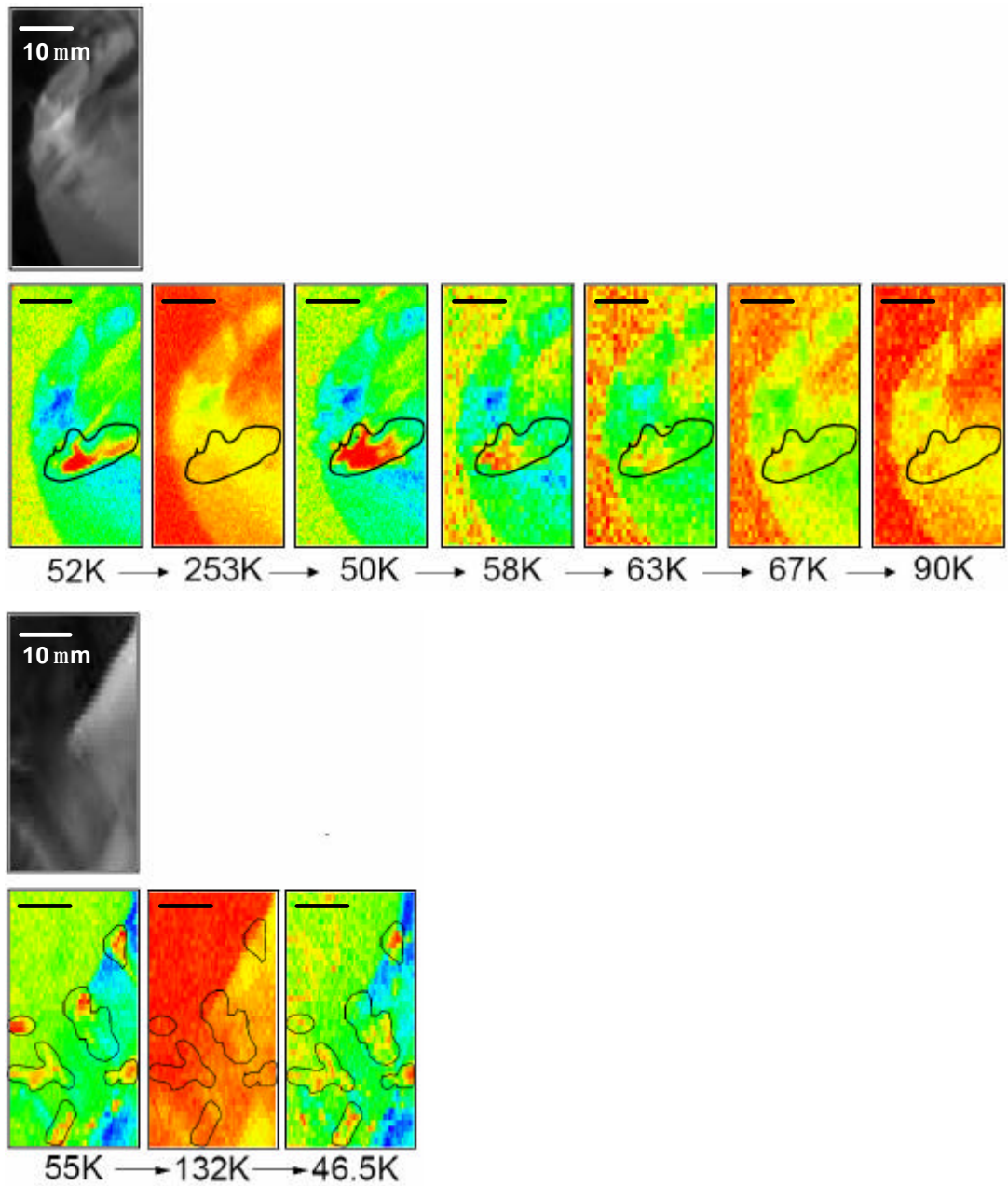

Fig 3 Sarma *et al*.